\documentclass[12pt,preprint]{aastex}

\newcommand{\ex}[1]{\mbox{$\times 10^{#1}$}}

\newcommand{\msol}{\mbox{$M_{\odot}$}}

\shorttitle{Coronographic Detections of Protoplanetary Disks}
\shortauthors{Karr et al.}
 
\begin{document}
\bibliographystyle{astron}

\title{Candidate Coronagraphic Detections of Protoplanetary Disks around Four Young Stars }

\author{J.L. Karr} \affil{Academia Sinica Institute of Astronomy and
Astrophysics, PO Box 23-141 Taipei 10617, Taiwan, ROC}
\author{N. Ohashi} \affil{Academia Sinica Institute of Astronomy and
Astrophysics, PO Box 23-141 Taipei 10617, Taiwan, ROC}
\author{T. Kudo} \affil{National Astronomical Observatory of Japan, 
2-21-1 Osawa, Mitaka, Tokyo 181-8588, Japan}
\author{M. Tamura} \affil{National Astronomical Observatory of Japan, 
2-21-1 Osawa, Mitaka, Tokyo 181-8588, Japan}

\begin{abstract}

We present potential detections of H-band scattered light emission
around four young star, selected from a total sample of 45 young stars
observed with the CIAO coronagraph of the Subaru telescope. Two CTTS,
CI Tau and DI Cep, and two WTTS, LkCa 14 and RXJ 0338.3+1020 were
detected. In all four cases, the extended emission is within the area
of the residual PSF halo, and is revealed only through careful data
reduction.  We compare the observed extended emission with simulations
of the scattered light emission, to evaluate the plausibility and
nature of the detected emission.

\end{abstract}
\keywords{stars:protoplanetary disks,stars:pre-main sequence}

\clearpage
\newpage

\section{Introduction \label{intro}}

A well developed picture of the formation of low mass stars has
emerged in recent years.  Stars form from the collapse of a dense,
dusty molecular envelope. As the material collapses, a circumstellar
disk forms around the protostar, through which material is accreted
onto the star. At the same time, an outflow propels material out of
the envelope. As the star evolves towards the main sequence, the
envelope first dissipates, then the optically thick circumstellar disk
becomes a thin disk and then a remnant debris disk, over a timescale
of approximately 10 Myr. While the general process is well
established, the detailed physical processes and time-scales are still
an area of active debate. The presence of circumstellar disks is
ubiquitous in this picture, and understanding their formation,
structure and evolution is a vital step towards understanding the
formation of both stars and the planetary systems that form in the
disks.

High resolution near-infrared (NIR) coronagraphic observations of
scattered light from circumstellar disks have revealed a complex and
intriguing variety of disks, ranging from the spiral arm structures of
AB~Aur \citep{fukagawa04}, to tail and ring-like structures around the
multiple systems T~Tau \citep{mayama06} and GG~Tau \citep{itoh02}, to
butterfly-like scattered light emission from edge-on thick disks in
Taurus \citep{padgett99} to the faint debris disks around the
relatively evolved sources TW~Hya \citep{krist00} and Beta~Pic
\citep{heap00}. A complimentary detection method of particular efficacy
in disk studies is the use of imaging polarimetry combined with
coronagraphy to observe linearly polarized scattered light from the
disk (eg. \citep{oppenheimer08, perrin04, apai04}); a similarly wide
variety of morphologies can be seen with this method as well.

Clearly, the complex variety of disks indicates that the processes
which shape circumstellar disks are much more complicated than
previously thought. There are many factors which can affect the
structure of the disk, including the luminosity and multiplicity of
the central star, the accretion characteristics, and the age of the
disk. In addition, the structure of the disk is intimately connected
to the formation of planets, as planetary systems can clear gaps in
the disks and the formation of planets can lead to the cessation of
accretion onto the star.

Directly resolving the surface of disks in scattered light is a
challenging task. In order to observe scattered light on solar system
scales, sub-arcsecond resolution is required for even the nearest star
forming regions; consequently, either spaced based observations are
required, or a ground based 8 m class telescope with adaptive optics
(AO). In addition, the scattered light emission is orders of magnitude
less than that of the central star. The stellar light needs to be
blocked using a coronagraphic mask, and great care taken in the
reduction and analysis of the data to avoid artifacts and false
detections.

Relatively few of the vast number of YSOs detected through infrared
photometric excess and spectroscopic features have been directly
imaged and resolved through infrared coronagraphy and polarimetry. A
much larger sample of objects is needed to build a consistent picture
for the evolution of disk structure. In this paper, we present
tentative detections from a large (45 object) sample of young stars
which have been observed using the Coronagraphic Imager with Adaptive
Optics (CIAO) on the Subaru 8.2 m telescope.

In Section~\ref{obs} we discuss the observation techniques and sample
for the survey. In Section~\ref{reduction} we discuss the details of
the data reduction procedure, and the technique used to choose the
most robust detections, weeding out instrumental artifacts. We detail
the detections and individual sources in Section~\ref{detect}, and
discuss the general results in Section~\ref{discuss}. In
Section~\ref{disks} we compare the spectral energy distributions
(SEDs) and scattered light emission to models and synthetic images, to
evaluate the nature and plausibility of the observed emission. The
results are summarized in Section~\ref{summ}.

\section{Observations \label{obs}}

\subsection{Coronagraphic Observations and Scattered Light}

The observations presented here are part of a large coronagraphic
survey of young stars with the Subaru telescope. This survey had two
goals; discovering low luminosity companions to young stars, and
directly resolving scattered light from disks in the
near-infrared. Near-infrared stellar light scattered from the surface
of a circumstellar disk can be a powerful tracer of the morphology and
extent of the circumstellar disk, although the interpretation of the
results is challenging. Unlike molecular line or sub-mm continuum
observations scattered light observations are sensitive to the surface
of the disk only, for an optically thick disk, rather than the total
amount of material. Consequently, these observations are potentially
sensitive to relatively low amounts of material. Conversely,
information about the interior of the disk is not probed.  An
additional complication in interpretation is caused by the fact that
scattered light can also be observed form the interior walls of the
outflow cavity of embedded young stars, and structures in the envelope
itself, leading to confusion regarding the source of the emission.

\subsection{The Sample \label{sample}}

All the imaging data were taken with the Subaru Coronagraphic Imager
with Adaptive Optics \citep{tamura00,murakawa04}, and were retrieved
from the public Subaru archive. The original, full sample of targets
was composed of roughly 100 targets selected for the likelihood of
observing a disk. The sample was not chosen as a statistically uniform
one; rather, it consists of a selection of sources of various masses
and spectral types which show strong evidence for the presence of a
circumstellar disk in a variety tracers.  Evidence for a circumstellar
disk can be inferred from excess emission in the near- to far-infrared
\citep{adams88}, resulting from the absorption and re-radiation of
stellar light by dust, from submillimeter and millimeter line and
continuum emission emitted by the molecular gas and dust, infrared
spectra showing dust features, and from X-ray emission (particularly
for weak-lined T-Tauri stars (WTTS)), H-$\alpha$ and lithium line
emission indicative of active accretion, which imply the presence of
an associated disk.

A large number of individual sources in the original sample have been
analyzed by a variety of researcher, particularly for the higher mass
Herbig Ae/Be stars and classical T-Tauri stars with optically thick,
more readily observable disks, and those sources which clearly show
evidence of extended structure in scattered light
(e.g. \citet{fukagawa04,mayama06,hioki07,itoh08,kudo08}). The
remaining sources, consisting primarily of lower mass K and M stars
and more evolved weak lined T-Tauri stars, do not immediately show
evidence for scattered light from disks. As a result, careful,
detailed data reduction and analysis are required in order to detect
extended emission from scattered light. The existence and nature of
circumstellar disks around WTTS is still a subject of much
debate. This makes the detection of disks around WTTS a less likely
proposition; at the same time it makes the detection of such a disk
scientifically very significant, and of great interest in furthering
the understanding of disk evolution. Therefore, the study of these
objects is of great interest in spite of the small amount of material.

In light of this, we have reduced and analyzed forty-five sources from
the above survey, in order to detect extended emission on small
physical scales ($<$ 200 AU). Table~\ref{full} lists all the sources
in this sample. This sample is heavily weighted towards lower mass
stars, particularly K and M stars, and weak-lined T-Tauri stars. The
sample is composed of 12 Classical T Tauri stars (CTTS), 31 weak line
T Tauri stars, 1 Herbig Be star and 1 YSO. The spectral types are
weighted towards lower mass stars; 8 M stars, 24 K stars, 7 G stars, 1
F star and three stars of unknown spectral type.  Most of the sources
are located in Taurus, at a distance of 140 pc, while a few are
located in other regions.  The complete 45 object sample, with
detection limits and companions, will be presented in a subsequent
paper.

Near-infrared coronagraphic observations in H-band were performed on
13 nights between November 2003 and November 2005.  The observing
times and number of frames used vary from source to source, and are
listed in Table~\ref{obsparm}. Even with the use of a coronagraphic
mask, there is still significant remnant flux in the wings of the
point spread function (PSF). Consequently, reference PSFs from
non-binary, non-disk sources were observed for each target, in order
to produce the near-complete post observation subtraction of the
PSF. Photometric standards from the UKIRT faint spectral standard
catalogue \citep{cat_faint} were observed on each night.

\section{Data Reduction \label{reduction}}

\subsection{Basic Reduction \label{basic}}

The data were reduced using the standard IRAF routines; dark frame
subtraction, flat fielding, bad pixel masking, cosmic ray reduction
and sky subtraction. The position of the source in the image can shift
slightly due to minor variations in the adaptive optics optics
performance during the night. The partial transparency of the CIAO
mask ($\approx$2\%) means that the central source is visible through the
mask. Consequently, the position of the central source can be used to
center each frame exactly.  The target frames were checked for AO
performance, using a measure of the AO quality which compares the peak
flux to the flux in the wings, measuring the sharpness of the image.
Each target and reference frame was visually inspected, to check for
bad frames, mis-aligned PSFs or other issues.

For each target source, the ten frames with the best adaptive optics
were selected. Ten PSF reference frames were then selected to best match
the quality of the adaptive optics in the target source.  The
reference frames were rotated to match the instrumental rotation of
each target frame, aligning the spider patterns on the image. The
reference frames were then co-added to create a reference PSF for each
individual target frame.

There are several possible methods for determining the best parameters
for the subtraction of the reference frame, which may need to be
shifted slightly in position (due to the centering of the source
within the mask), and/or scaled to match the intensity of the target
frame. We have adopted a two phase approach; an automated routine,
followed by a manual fine-tuning.  In the first phase, the averaged
intensity in the wings of the PSF was used to estimate the scaling
factor. The shift in the position of the central pixel was determined
by calculating the noise in the wings of the PSF and minimizing the
difference between the noise at four positions located away from the
diffraction spikes.  In the second phase, we interactively checked the
parameters derived above, to test for the quality of the fit and the
robustness of the parameters. It should be noted that the automated
algorithm will work best in the absence of any extended emission, and
will result in over-subtraction for symmetric extended emission.  At
this point, corrections to the position and scaling were generally
minor, typically less than a few pixel in position and 10\% in
scaling. The final subtracted images were then co-added to produce a
final image.

PSF subtraction for binary sources is a more complex process. In this
case, a reference PSF was created by adding two PSFs, each scaled and
shifted to match the relative fluxes an positions of the binary
components, before subtraction. 

\subsection{Detecting Extended Emission \label{extend}}

Potential detections were selected through a multi-stage process. Only
one source, Haro 6-10, showed clear extended emission in the original
coronagraphic image, as is the typical case in coronagraphic
detections of extended emission from disks and outflows.  Haro 6-10 is
a known embedded binary protostar, and the detected emission is not
likely to be disk emission, but rather scattered light off the
interior of the outflow cavity \citep{koresko99}. For the remaining
sources, potential detections were limited to the region within the
radius of the PSF halo.  As a consequence, the quality of the final
subtraction is in large part determined by the fine details of the AO
quality in the images, and the matching of the target and reference
PSFs prior to subtraction. Subtraction artifacts are therefore of
major concern.

 The detectability of scattered light from the surface of disks
 depends on the brightness of the scattered light, the orientation and
 morphology of the disk, and the distance of the source.  The
 coronagraphic observations are also highly dependent on the technical
 details of the observations.  The coronagraphic mask size determines
 the minimum angular scale at which scattered light can be detected;
 this is in practice larger than the mask size. The quality of the
 data is also highly sensitive to the performance of the adaptive
 optics system and the subtraction of the residual PSF effects. Even
 with the selection of the best quality image frames there are
 noticeable variations in the size of the PSF. Poor AO performance and
 mis-matched PSFs can either mask existing emission or create false
 detections.  For sources which show extended emission outside of the
 area dominated by the PSF halo, as in most previously reported cases,
 this makes the selection of the best PSF subtraction parameters
 problematic. For extended emission that is contained entirely within
 the halo dominated region, as in the sample presented here, the
 uncertainty comes in at the detection level, rather than the
 interpretation level.

\begin{figure}
\plotone{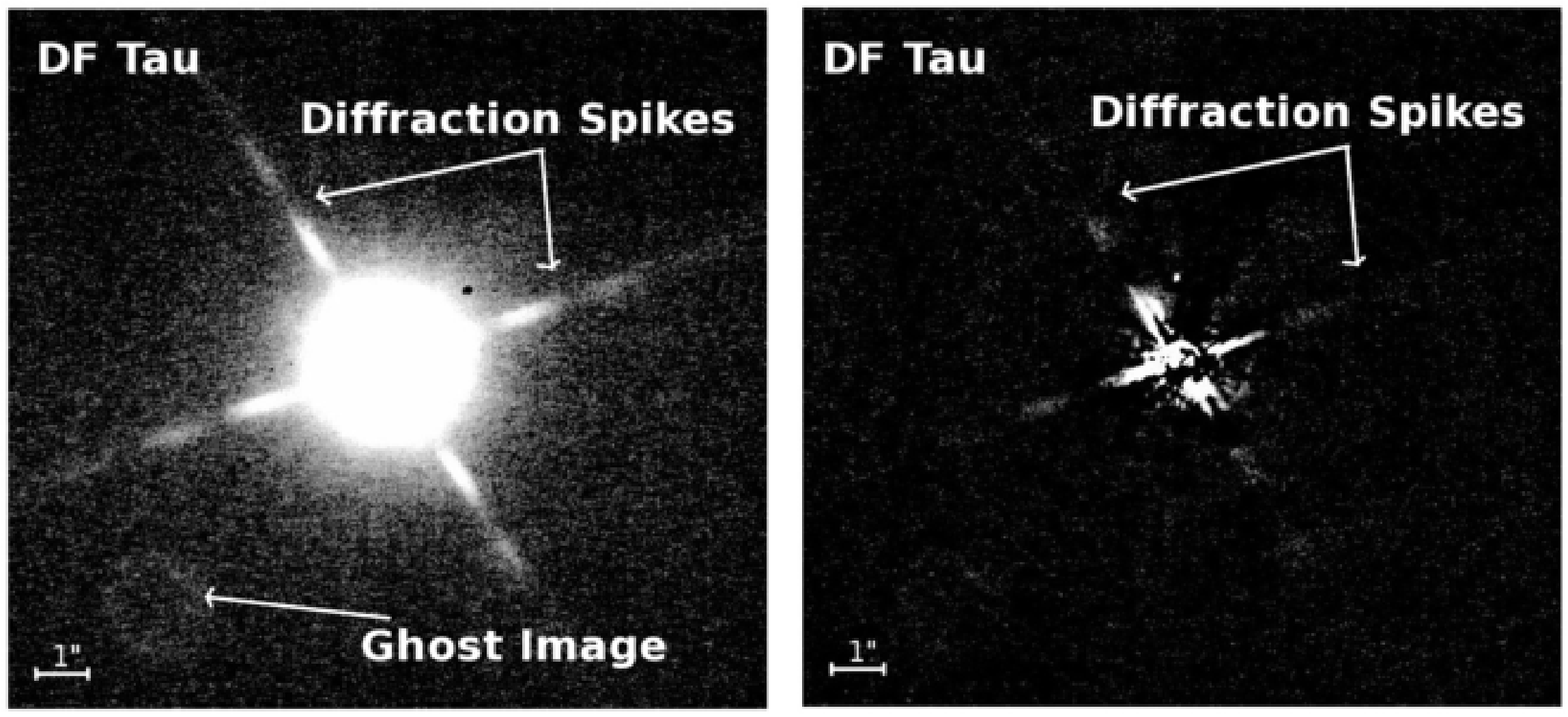}
\caption{Example of a null detection (no extended emission) for
  DF~Tau.  The left panel shows the pre PSF subtraction image, with
  instrumental artifacts labeled. The right panel shows the post
  subtraction artifacts.  Compare with the potential detections of
  Figure~\ref{all}.
\label{null}}
\end{figure}

\begin{figure}
\epsscale{0.75}\plotone{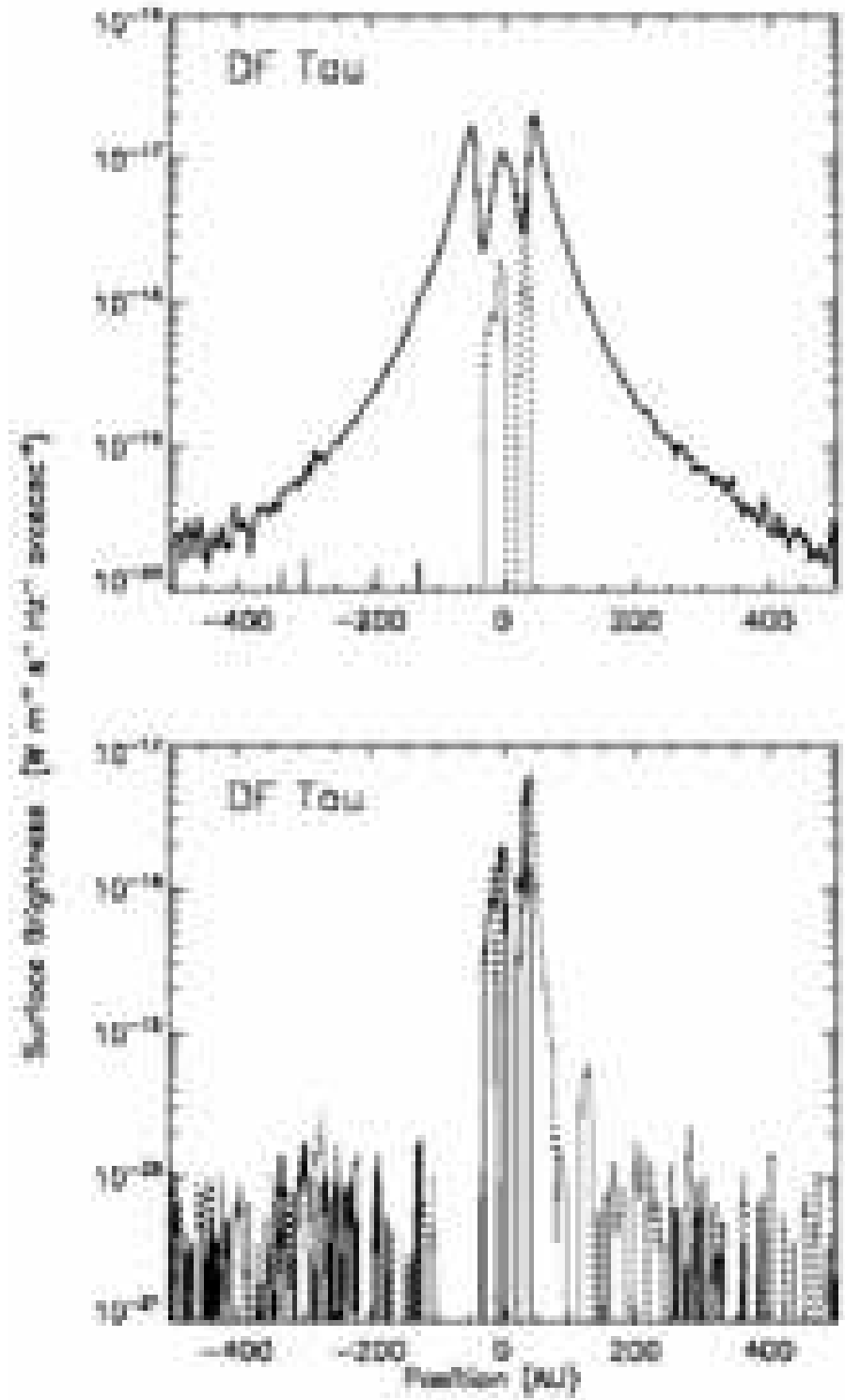}
\caption{Profiles for the non-detection of Figure~\ref{null}. The top
  panel shows pre-(solid) and post (dashed) subtraction profiles along
  an axis off of the diffraction artifacts. The lower panel shows the
  two post subtraction profiles, perpendicular to each other.  Compare
  with the detections in Figure~\ref{profiles}.  In all cases the
  profiles are averaged over seven pixels. \label{unprofiles}}
\end{figure}

A source was rejected if it showed a good but negative PSF
subtraction, in which there was no residual extended emission or
remaining artifacts.  A source was also rejected if it showed a poor
but negative PSF subtraction, with no remnant extended emission but
significant artifacts from the subtraction process. Figure~\ref{null}
shows an example of a rejected subtraction; DF~Tau is the result of a
null result from a moderately bright source.  The inner region of the
subtraction, immediately outside the mask, is dominated by subtraction
artifacts and is useless for the purposes of data analysis, even in
example of clear detections.  In practice, subtraction artifacts from
the diffraction spikes remain even in the best subtractions and for
bright sources in particular, these are impossible to remove
completely.  In Figure~\ref{unprofiles} we show the pre and post-PSF
subtraction profile for the same source, more quantitatively
illustrating the magnitude of the halo subtraction effect; compare
with the detection profiles discussed in Section~\ref{detect}.  If a
source showed evidence of extended emission, it was rejected unless
the extended emission was present with the same morphology and
direction in all ten of the individual frames.

We performed a series of tests on potential detections to attempt to
rule out instrumental or subtraction artifacts. We checked the
variation of the PSF from frame to frame by eye; a visible variation
in PSF size was noticeable in most observation runs, but a distortion
in the shape of the PSF was not readily apparent. A rotation of the
reference source by 90 degrees before subtraction was in general
inconclusive; small scale emission was dominated by the effects of the
diffraction spikes.  When possible, we tested variations in the
individual reductions to check the robustness of the subtraction.

In addition, we checked for systematic effects in the observational
data and processing which could explain the detected sources. We
compared the relative rotations of the target and reference, the
brightness of the sources, and the quality of the PSF matching between
the target and reference source. The potential detections did not show
any systematic biases when compared to the rest of the sample, in
relative rotation angle, flux of the source, quality of the PSF
matching or night observed.

The profile of the extended emission was calculated, and compared to
the profile from known bad subtraction artifacts. This was also
relatively inconclusive, due to the small angular scale over which
extended emission was observed.

œôøË
\section{Results \label{detect}}

At the end of the above process we retained four potential detections
(excluding Haro 6-10). Each detection, in itself, is marginal; it is
the large number of comparison non-detections in this sample which
indicates that they are of interest and worthy of subsequent,
confirming observations.  There are two CTTS, CI Tau (located in
Taurus) and DI Cep (at a distance of 300 pc), and two WTTS, LkCa 14
and RXJ 0338.3+1020, both in Taurus.  CI Tau, DI Cep and LkCa 14 all
show nebulous emission that is extended along one axis, while RXJ
0338.3+1020 shows more symmetric emission.

\begin{figure}
\epsscale{0.7} \plotone{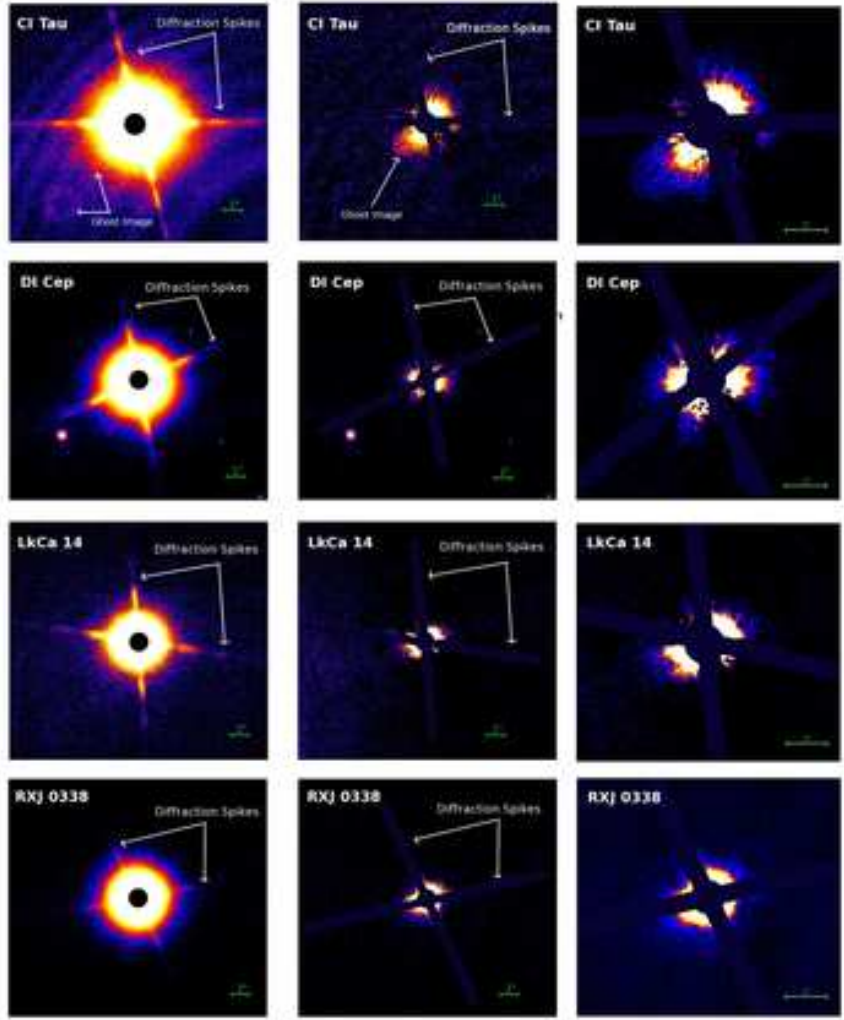}
\caption{Pre-(left) and post (middle,right) subtraction images for all
  four potential detections. The diffraction patterns are marked, and
  masked out in the subtracted image; as is the central, artifact
  dominated region. The ghost image near CI Tau is also indicated. The
  left and middle panels are on the same physical scale, the right
  panel shows a close-up of the extended emission.
\label{all}}
\end{figure}

\begin{figure}
\plotone{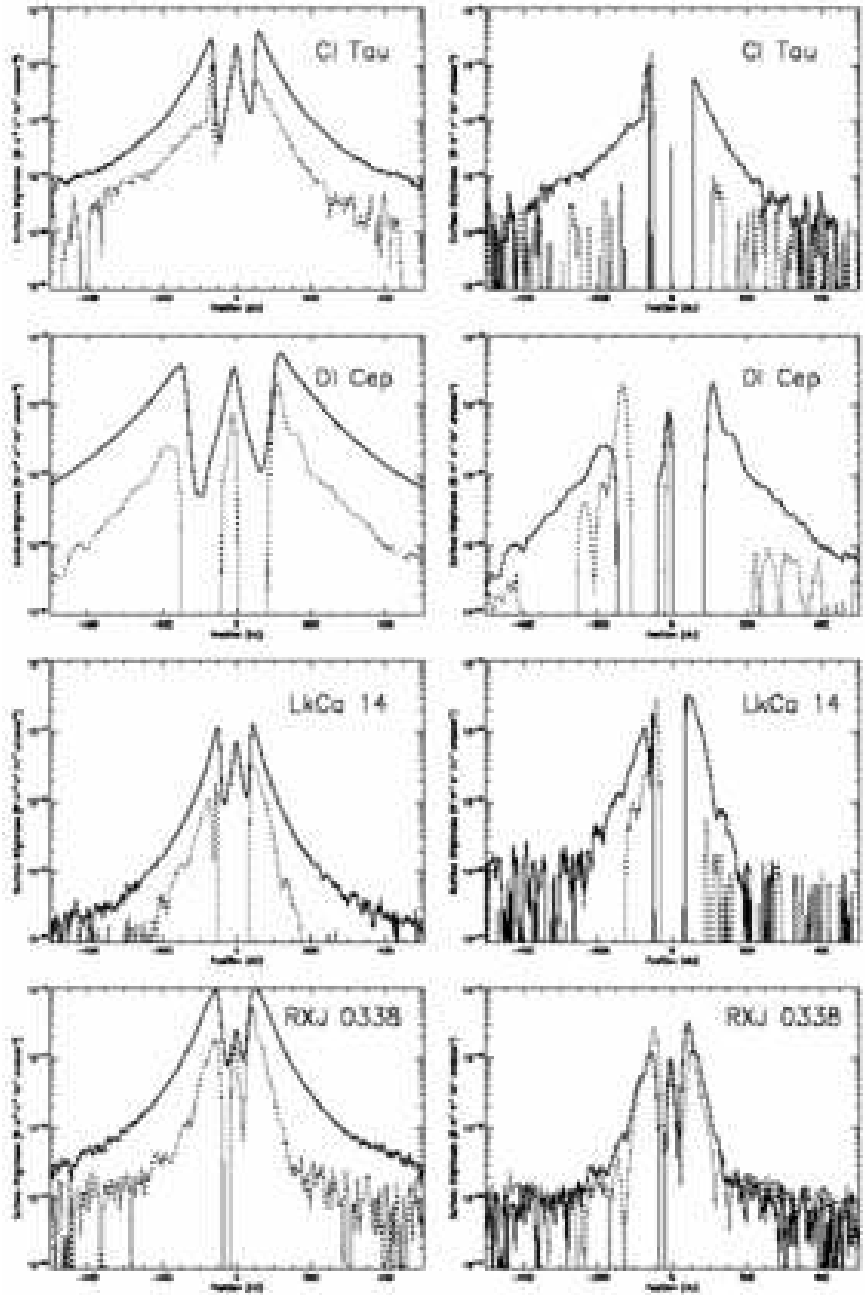}
\caption{Pre-and post PSF subtraction profiles all four potential
  detections. The left panels show the pre-(solid) and post (dashed)
  subtraction profiles along the major axis of emission, the right
  panels show the post subtraction profiles along (solid) and
  perpendicular to (dashed) the major axis of emission. In all cases
  the profiles are averaged over seven pixels.
\label{profiles}}
\end{figure}

Table~\ref{source} lists the physical of the potential detections;
position, spectral type, YSO class, distance, binarity and presence of
a companion in the field of view, detection limits of the
observations, and morphology and scale of the extended emission.
Figure~\ref{all} shows pre- and post-subtraction images for the
detections. The central region, dominated by subtraction artifacts,
has been manually masked. In the subtracted image the region dominated
by the diffraction spikes has also been masked for interpretive
purposes.  Figure~\ref{profiles} shows the corresponding profiles. In
the left panel, the pre- and post-subtraction profiles are compared,
while in the right panel the profiles along the major axis of emission
and perpendicular to it are shown.  In all cases the profiles are
averaged over seven pixels in width.  When compared with the
non-detection profiles of Figure~\ref{unprofiles}, it can be seen that
the the extended emission is much stronger and more extended along the
major axis of the detections than along the minor axis or the
non-detections. The magnitude of the residual PSF emission, compared
to the detected emission, is also clearly seen, and the need for
careful PSF subtraction.

\subsection{CI Tau}

CI Tau is a classical T-Tauri star of type K7, located in the Taurus
region at a distance of 140 pc \citep{kenyon95}. The post-subtraction
image of CI Tau in Figure~\ref{all} shows a nebulosity extending from
NE to SW.  To the south-west there is some contamination from a ghost
image in the target frame, which has been labeled.  The physical
extent of the nebulosity is approximately 190 AU, and the position
angle 145 degrees east of north.

There have been previous resolved detections of the circumstellar disk
around CI Tau. \citet{dutrey96}, in 2.7 mm continuum observations with
the Plateau de Bure interferometer, made a marginal resolution of the
disk, with an extent of 2.0 by 0.8 arcseconds and a position angle of
40 degrees. Later interferometric sub-mm observations performed by
\citet{andrews07} at 340 GHz with the SMA resolved emission. They fit
a disk with a radius of 225 AU, a mass of 0.04 $\msol$, an
inclination of 46 degrees and a position angle of 131 degrees, nearly
perpendicular to the previous detection. It should be noted that these
observations were barely resolved.  The latter angle is consistent
with the major axis of our NIR scattered disk, in which case the
scattered light images might show the light from the surface of the
disk detected with SMA.

\subsection{DI Cep}

DI Cepheus is a CTTS with a spectral type of G8, located at a distance
of 300 pc \citep{gameiro06}. We detect a nebulosity, shown in the
second two panels of Figure~\ref{all}, that is similar in morphology
to CI Tau, although the greater distance gives a physical scale of 370
AU for DI Cep. The position angle is 95 degrees. A companion candidate
can be seen to the south east of the image.

There have been no high resolution observations at the scales probed
by the Subaru observations. The strong infrared excess in the SED
(discussed further in Section~\ref{seds1}), is strongly indicative of
extensive circumstellar material, while the H$\alpha$ line width
\citep{ hessman97} implies accretion, which assumes the presence of a
disk. Mid infrared spectroscopy by \citet{bowey03} detected strong
silicate emission from circumstellar dust; they fit a disk with a
10$^4$ AU radius to the line profile.  \citet{gameiro06} performed
high resolution optical spectroscopy of DI Cep, and argue that its
disk is observed nearly edge on. The morphology of the scattered light
images indicate a disk that is seen at an angle from face on, although
not likely directly edge on.

\subsection{LkCa 14}

LkCa~14 (aka V1115 Tau) is a weak-lined T-Tauri source with a spectral
type of M0V, also located in the Taurus region \citep{kenyon95}. The
extended emission shown in the third set of panels in Figure~\ref{all}
is of smaller scale that the other sources, and as a result it is the
weakest of our detections. The emission is again elongated,
with a physical radius of 120 AU and a position angle of 135
degrees. The corresponding profiles are shown in the third panel of
Figure~\ref{profiles}.

There has been no previous, resolved detection of the disk around
LkCa~14.  \citet{andrews05} place an upper limit on the mass of the
disk of 0.004 $\msol$, based on single dish sub-mm observations at 340
GHz. H$\alpha$ is seen in emission, which is an indication of
accretion, and indirectly of the presence of circumstellar material.
It should be noted that coronagraphic observations probe scattered
light off the surface of the disk, and consequently a relatively small
amount of material is needed to produce a scattered light image.

\subsection{RXJ 0338}

RX J0338 is a weak-line T-Tauri star of type G9, also located in
Taurus at 140 pc \citep{magazzu97}.  It shows irregular extended emission, as
seen in the last two panels of Figure~\ref{all}, with a radius of 190
AU.  Very little is known about this X-ray selected source.

\section{General Discussion \label{discuss}}

The detections reported above are tenuous ones; it is only the large
scale nature of the survey and the comparison with multiple
non-detections that allows us to select these as interesting sources
for follow-up observations.  One clear result from the analysis of
these data is the strong dependence of the results on the quality of
the adaptive optics performance.  Consequently, these sources are
likely best viewed as the most promising candidates for subsequent
observations, rather than as firm detections.  Further observations
with more advanced coronagraphic and adaptive optics (particularly
HICIAO and the new 188 element adaptive optics system at Subaru
\citep{hodapp08}) are in the planning stages, while observations at
longer wavelengths could probe the inner disk material, and ideally
provide velocity information to determine the disk orientation and
provide an independent comparison to the infrared morphology.

Of the four sources shown here, three show elongated structures, while
one shows more circularly symmetric structure. In addition to light
scattered from the surface of the disk, light may also be scattered
from the interior walls of an outflow cavity; with single wavelength
data it can be difficult, if not impossible, to distinguish between
light from a small scale outflow cavity and an inclined disk. Further
data are needed. However, it should be observed that of the three
elongated features, one is a weak line T-Tauri star, which should
not show any significant outflow activity. Even the classical T-Tauri
stars should have a relatively minor envelope compared to the earlier
phases of star formation; this is supported by the optical visibility
of the sources.  The emission is also relatively symmetric about
the central star. In the case of an outflow cavity, sources are rarely
observed face on, and therefore the red-shifted lobe is generally more
highly extincted than the blue-shifted one, due to the geometry of the
system. Both of these observations argue against an outflow cavity
interpretation for these objects.

The detections around the WTTS are, if confirmed, of great
interest. The nature and extent (or even presence) of disks around the
more evolved WTTS are not well constrained, and the small amounts of
material involved make observations in the mm or sub-mm
difficult. Consequently, a positive detection of a disk around a WTTS
will have very interesting implications for the long term evolution of
circumstellar disks.

The disk properties and frequency of WTTS are not well constrained,
either by observation or by theory. The disks themselves are low mass
and consequently low luminosity. From a theoretical perspective, the
disk dissipation timescale is of importance, and also not well
constrained.  Ground based photometry is not sufficient for the
detection of WTTS disks in the infrared, and earlier space missions
such as ISO and IRAS were not of sufficient sensitivity, due to the
small excesses at longer wavelengths and the photospheric SEDs at
shorter wavelengths.  Post Spitzer statistics for infrared derived
WTTS disk fractions vary considerably, from 30-35\% for WTTS in star
forming clusters \citep{lada06,sicilia-aguilar06,padgett06}, to 23\%
for WTTS in the Taurus region \citep{cieza07}, to 6\% for the areas
surrounding star forming regions \citep{padgett06}.  Sub-mm studies of
well known sources have derived disk fractions for WTTS (with masses
in the range of 10$^{-1}$ to 10$^{-3}$ $\msol$) of
10\% \citep{osterloh95, andrews05}.

From these recent studies it is clear that a substantial minority of
WTTS have remnant circumstellar disks. This is an encouraging result
for further, more sensitive, coronagraphic studies, as a relatively
small amount of material is necessary to produced scattered light
images. One area of uncertainty, however, is the scattering properties
of the dust around WTTS.  For a WTTS, planetary formation and the
fragmentation of planetisimals will strongly shape the dust
properties. Even at earlier phases, dust settling and coalescence are
also expected to alter the scattering and emission properties of the
disk.  A WTTS disk with significantly processed dust, and consequently
much larger average grain size, might result in a significantly
fainter scattered light image for a given disk mass. Further, longer
wavelength observations are needed for the sources in this paper to
further constrain their disk properties.

\section{Constraining the Disks \label{disks}}

Given the tenuous nature of the above detections, it is important to
look at the plausibility of the detections from the perspective of the
physical size and mass of the potential disks. There are several ways
this question can be approached. The photometry of the sources can be
used to explore the range of potential disk parameters which are
consistent with the shape of the SED.  Scattered light images
generated from radiative transfer models can be used to probe the
nature of the scattered light emission for different disk
configurations.

\subsection{SEDs and Disk Parameters \label{seds1}}

A team associated with the Spitzer GLIMPSE Legacy survey
\citep{robitaille06, robitaille07} has developed a powerful online
tool for fitting SEDs to photometric observations of YSOs, making it
possible to efficiently analyze SEDs in the context of physical models
of YSO systems. Fitting detailed YSO models to an SED using radiative
transfer simulations is a non-trivial task. The group takes a novel
approach; first generating a suit of 20,000 3-D radiative transfer
models \citep{whitney03a,whitney03b} for a grid of 2-D YSO models,
using a Monte-Carlo technique which traces the path of individual
photons.  The models cover a wide range of stellar masses and ages,
 masses and geometries for disks, envelopes and outflow cavities,
and a set of ten inclinations for each model. Observed SEDs, from the
optical to millimeter, are then fitted to the grid of models, with the
addition of distance and interstellar extinction effects.

There are several important caveats to this method of fitting. The
most important is the degeneracy of the fits. With the large number of
physical parameters input to the radiative transfer model, the fits
are highly non-unique, with multiple combinations of parameters
fitting a given SED.  Consequently, care must be taken not to
over-interpret the physical significance of the results.  However, the
fits can be used to interpret the plausibility of a given set of
models, and to constrain the general best fit of given a
parameter. This works better for some parameters than others; the disk
inclination, in particular, is poorly constrained by the SED fits.

\begin{figure}
\plotone{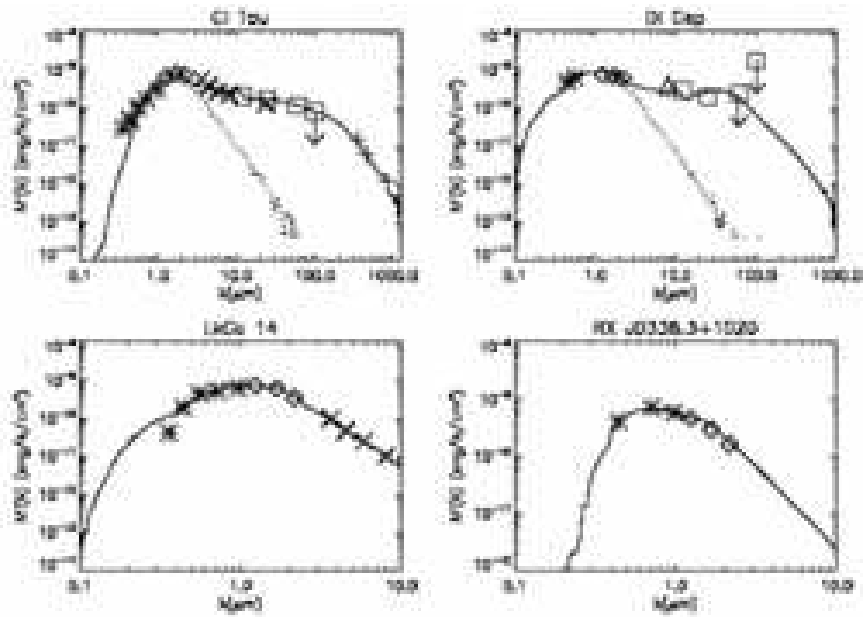}
\caption{SEDs for the tentative detection; Stars; optical photometry
  (U-I band); Diamonds; 2MASS JHK$_s$ photometry; Crosses; Spitzer;
  Squares; IRAS, Pluses; sub-mm/mm data. Upper limits are indicated by
  downwards arrows.  The solid line is the SED from the disk model
  \citep{robitaille06},the dashed line the component from the stellar
  black-body only, for the WTTS these are identical.
\label{seds}}
\end{figure}

Figure~\ref{seds} shows the SEDs and fits for the tentative
detections.  In each case, we have used the best literature photometry
available (optical, near- to far-infrared and sub-mm/mm); the
photometry is summarized in Table~\ref{phot}.  The distance and
interstellar extinction have been constrained to measured values in
the initial fit.  From the suite of models which result, we have then
selected the best model which matches the stellar mass of our source.

CI~Tau and DI~Cep both show strong infrared excesses in the
mid-infrared and beyond. CI~Tau is fit with a 0.4 $\msol$ star with a
0.03 $\msol$ disk, similar to observational estimates of the disk
mass, while DI~Cep is fit with a 1.1 $\msol$ central star, with a disk
mass of 0.03 $\msol$.

LkCa~14 and RXJ~0338, on the other hand, are adequately fitted by an
extincted stellar spectrum in the range of the available photometry.
This is consistent with their classifications as WTTS, where little,
if any, infrared excess is generally seen. They do not, therefore,
have a large disk, but the SEDs do not rule out the possibility of a
low mass remnant disk, or a cooler disk at larger radii with a
central gap. The latter would show an excess only at longer
wavelengths.  The model for LkCa~14 shown is for a 0.4 $\msol$ star
with a disk of 2.1$\ex{-4}$ $\msol$, within the observational limits,
while RXJ shows a photospheric SED for a central star of 5600 K, the
best fit stellar spectrum.

\subsection{Modeling the Extended Emission}

For further analysis of the morphology of the NIR emission, the
scattered light properties of the above SED models can be determined.
The radiative transfer codes used for generating the suite of models
may be downloaded by the user and run directly. This generates
multi-wavelength images that can be directly compared with high
resolution observations, after convolution with the relevant
instrument parameters.

For the purposes of this paper, we have used the SEDs fit using this
technique, as discussed previously in Section~\ref{seds1}, as input to
the radiative transfer code to generate high resolution images. It is
important to note that this is not a fit to the images; rather this
allows us to explore how probable our observed images are given the
source SED and reasonable system geometries. Several appropriate disk
models were tested, to explore the range of emission parameters for
the given SED. The inclinations, which are poorly constrained by the
SEDs but strongly affect the morphology of the scattered light
emission, were tuned to match the morphology of the extended emission.

\begin{figure}
\plotone{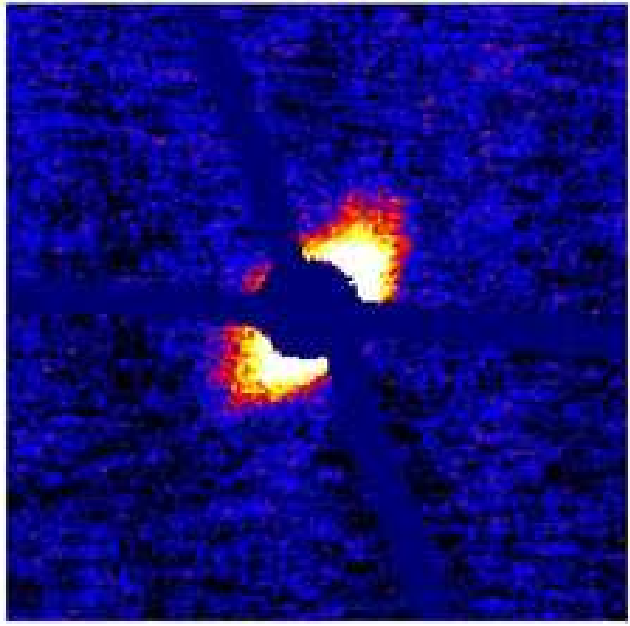}
\caption{Simulated image for a disk which is consistent with the SED
  of CI Tau, for a disk with a mass of 0.03 $\msol$ and outer radius of
  133 AU, inclination of 63 degrees.
\label{model1}}
\end{figure}

The results for CI~Tau are shown in Figure~\ref{model1} for a model
with a total disk mass of 0.03 $\msol$, outer radius of 133 AU and a
disk inclination of 63 degrees, matching the observed extinction and
distance of the source. To compare with the observed data, a mask has
been manually added to the images where the coronagraphic mask and
diffraction spikes would dominate the image in real data, and the
central few pixels (containing the stellar flux) have been removed.
The simulated image has then been convolved with an approximation of
the Subaru PSF, measured within the coronagraphic mask, and background
noise added to match the final co-added image.  The extent, morphology
and surface brightness of the observed extended emission can be
reproduced with a physical disk/star model that is consistent with the
observed SED.  The flux levels match approximately, although the model
levels are lower (by $~$50\%) than the observed values. The physical
scales are similar; however the emission in the model drops off
abruptly when the edge of the disk is reached, while the observed
emission tapers off more gradually.

\begin{figure}
\plotone{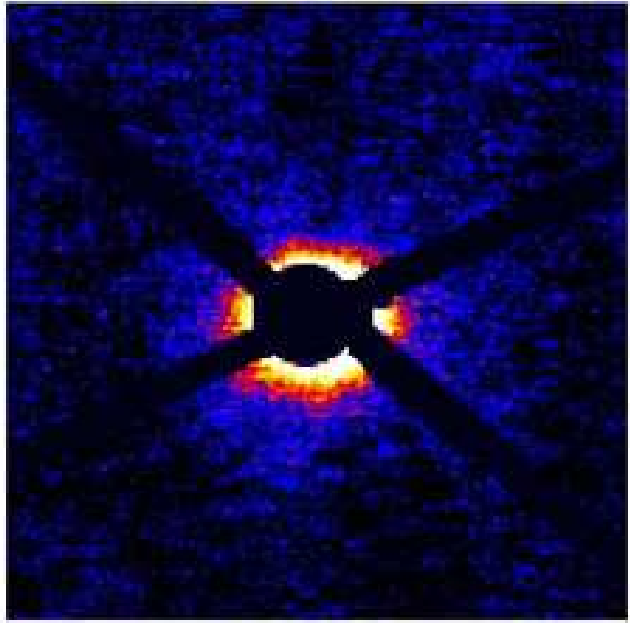}
\caption{Simulated image for a disk which is consistent with the SED
  of RXJ~0338, for a disk with a mass of 2.8$\ex{-3}$ $\msol$, with an
  inclination of 44 degrees and an inner disk gap of 20 AU radius. The
  inner disk gap is within the mask, and not visible.
\label{model2}}
\end{figure}

For the WTTS, the situation is more difficult, as the observed SED
only covers the stellar photosphere portion of the spectrum, with no
observed excess emission.  Therefore, fitting the disk portion of the
SED is not possible.  However, there are some limits on the parameters
of the disk, as an infrared excess is not seen at NIR wavelengths. For
RXJ~0338, we ran the SED fitter to produce a suite of consistent
models; typically low mass disks with a central hole. We selected
several SED models which matched the central source, and calculated
the corresponding scattered light images. Figure~\ref{model2} shows an
image for a disk which is consistent with the observed SED. The model
has a disk mass of 2.8$\ex{-3}$ $\msol$, with an inner disk gap of 20
AU and an inclination of 41 degrees.

One area of uncertainty in the calculation of the scattered light
images is the dust emission characteristics, as the radiative transfer
code does not accurately model the dust properties expected for a
highly evolved disk.  The original models used in the SED fits were
generated using dust emission properties derived from the ISM
\citep{whitney03a}.  Calculating the scattered light properties with
the same disk geometry and mass, but using the alternate dust models
provided with the code \citep{cotera01,wood02, whittet01} does produce
a noticeable difference in the scattered light properties, with a
variation of about 50\% in the size of the disk, measured at a given
surface brightness level.

\section{Summary \label{summ}}

We have presented four potential detections from a large (45 object)
coronagraphic study of YSOs observed with the Subaru telescope. The
detections are marginal, but have been extensively tested for the
robustness of the detection against the non-detections in the the
sample. 

Two of the detections, CI~Tau and DI~Cep, are CTTS with extensive
infrared excesses; the disk around CI~Tau has been previous resolved
in the sub-mm. Both sources show elongated emission indicative of an
inclined disk, CI~Tau with a physical radius of 190 AU, the more
distant DI~Cep with a physical radius of 370 AU.  The remaining two
detections, RXJ~0338 and LkCa~14, are WTTS and have no previous direct
detection of the disks. RXJ~0338 shows roughly symmetrical emission
with a radius of 190 AU, while LkCa~14 shows elongated emission with a
radius of 120 AU. The WTTS detections, if confirmed, are of great
interest, as the nature and frequency of disks around WTTS are not
well constrained.

We have fit the SEDs of the sources using the SED fitter of
\citet{robitaille06}, and used several of the resulting fits as input
to the companion radiative transfer code.  The general morphology and
brightness of the scattered light emission from the CTTS can be
reproduced with a radiative transfer model of a disk system that is
consistent with the optical to mm SED. The calculation of scattered
light emission for these sources is harder to constrain, as the
observed SEDs do not extend to long enough wavelengths to detect the
infrared excess. However, the general morphology of the emission can
be reproduced using a disk model that does not contradict the observed
portion of the SED.

Further observations are needed to confirm these detections, in
particular future coronagraphic studies. For the WTTS, detections of
the infrared excess at longer wavelengths will provide a constraint on
the nature of the surrounding disk. 

\newpage

\section{Appendix A: Reduction Details}

The following section contains details of the reduction process
primarily of interest to readers with experience in the processing of
coronagraphic data. 

\begin{figure}
\plotone{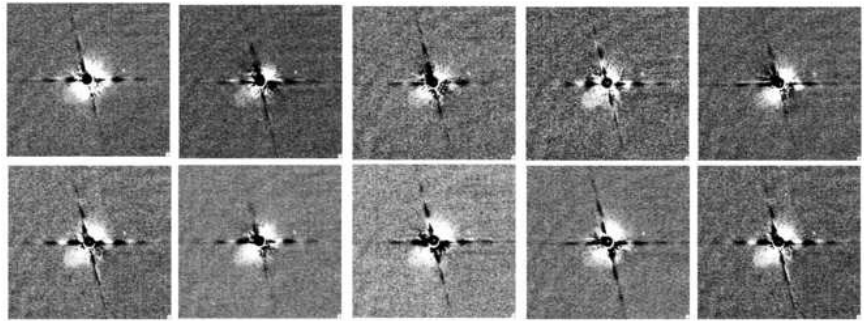}
\caption{Details of the reduction of CI Tau, showing the results of individual frames. 
\label{good}}
\end{figure}

\begin{figure}
\plotone{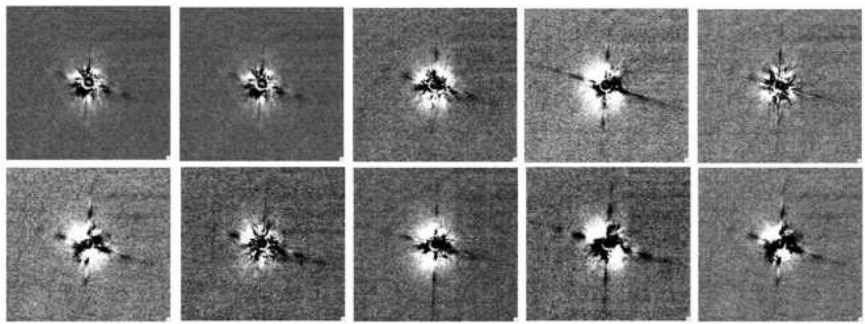}
\caption{Details of a messy null detection which fails the selection
  process (CW Tau)
\label{bad}}
\end{figure}

Figures~\ref{good} and \ref{bad} shows details of the source rejection
pipeline for a source which passes the tests, and for a source which
fails the process on several counts. Each sub-image shows the result
of a single subtraction, before co-adding. For all the sources, the
final subtractions were inspected by eye, in sequence.  The final
images shown in this paper are the coadded results of the individual
frames, with diffraction spikes and central region manually masked.

Figure~\ref{good} shows the details of the data reduction for CI
Tau. This source passes the detection as extended emission appears in
all ten frames, with the morphology and extend similar in all frames;
note that there are still some variations in brightness resulting from
the subtraction process. In the case of randomly fluctuations in the
atmosphere or performance of the adaptive optics, such a coherence in
the morphology from frame to frame would not be expected. 

Figure~\ref{bad} shows a source which was initially flagged as
possibly showing extended emission, but which later fails the sequence
test. Extended emission is not present in half of the best frames, and
sources which do show extended  emission vary in mophology from frame
to frame.  Therefore the source is rejected as a likely residual.

\newpage

\clearpage
\newpage

\begin{deluxetable}{lll|lll}
\tabletypesize{\scriptsize} 
\tablecaption{The Entire Sample \label{full}} 
\tablewidth{0pt} 
\tablehead{\colhead{ Name } & \colhead{ Spectral Type} & \colhead{ Evolutionary Status} 
&  \colhead{ Name } & \colhead{ Spectral Type} & \colhead{ Evolutionary Status}} 

\startdata 
CI Tau            & GV:e       &CTTS          &    RX J0423.5+0955   & K4         &WTTS          \\       
CW Tau            & K5V:e      &CTTS          &    RX J0426.4+0957W  &            &WTTS          \\      
DE Tau            & M1V:e      &CTTS          &    RX J0438.2+2023   & G9         &WTTS          \\      
DF Tau            & K5         &CTTS          &    RX J0445.3+0914   & G0         &WTTS          \\      
DP Tau            & M0V:e      &CTTS          &    RX J0512.0+1020   & K2         &WTTS          \\      
DS Tau            & K4V:e      &CTTS          &    RX J0528.4+1213   & K2         &WTTS          \\      
FM Tau            & K3         &CTTS          &    LkCa 1            & M4V        &WTTS          \\      
GI Tau            & K5e        &CTTS          &    LkCa 4            & K7:V       &WTTS          \\      
HO Tau            &            &CTTS          &    LkCa 5            & M2V        &WTTS          \\      
ZZ Tau IRS        & M5.2v      &CTTS          &    LkCa 14           & M0:V       &WTTS          \\      
HBC 347           & K1         &WTTS          &    LkCa 19           & K0V        &WTTS          \\      
HBC 366           & M0V        &WTTS          &    HD 17543c         & F8V        &WTTS          \\      
HBC 374           & K7         &WTTS          &    HD 41593          & K0         &WTTS          \\      
HBC 376           & K7         &WTTS          &    V386 Cep          & S          &WTTS          \\      
HBC 388           & K1         &WTTS          &    Haro 6-10         & K3         &Class I       \\      
HBC 392           & K5         &WTTS          &    BD+31 643         & B5         &Herbig Be     \\      
HBC 397           & K7         &WTTS          &    HDE 283572        & G2III      &WTTS          \\      
HBC 403           & K7         &WTTS          &    DI Cep            & G8V        &CTTS          \\      
HBC 415           & Gn         &WTTS          &    BM And            & K5V        &CTTS          \\          
RX J0336.0+0846   & M3         &WTTS          &    V826 Tau          & K7Ve       &WTTS          \\          
RX J0338.3+1020   & G9         &WTTS          &    V827 Tau          & K7         &WTTS          \\          
RX J0404.4+0518   & K0         &WTTS          &    V830 Tau          & K7         &WTTS          \\          
RX J0409.8+2446   & M1.5       &WTTS          &

\enddata
\end{deluxetable}

\begin{deluxetable}{llllll}
\tabletypesize{\scriptsize} \tablecaption{Observational Parameters\label{obsparm}} 
\tablewidth{0pt} \tablehead{\colhead{Source}  & \colhead{Date Observed} & \colhead{Photometric Reference$^{1}$} & \colhead{Number of Frames Observed} & \colhead{Exposure Time/Frame (s)} & \colhead{PSF Reference} }
\startdata 
RX J0338.8+1020   & 2003 Dec 3  & FS 120           &  117     &    10 &  SAO 76545  \\
SAO 76545     &     2003 Dec 3  & FS 120           &  144     &    10 &  		  \\
CI TAU        &     2005 Nov 10  & FS 125           &  100     &    15 &  SAO 76560  \\
SAO 76560     &     2005 Nov 10  & FS 125           &  170     &    12 &  		  \\
LkCa 14       &     2003 Nov 4  & FS 115/FS 4      &   36     &    30 &  HD 286794  \\
HD 286794     &     2003 Nov 4  & FS 115/FS 4      &   45     &    30 &  		  \\
DI Cep        &     2005 Nov 12  & FS 125           &   60     &    20 &  SAO 34978  \\
SAO 34978     &     2005 Nov 12  & FS 125           &   96     &    10 &             \\
\enddata
\tablenotetext{1}{UKIRT Faint Spectral Standards \citep{cat_faint}}
\end{deluxetable}

\begin{deluxetable}{llllllllllllllllllllllllll}
\tabletypesize{\tiny} \rotate \tablecaption{Sources with Potential Detections \label{source}} 
\tablewidth{0pt} \tablehead{\colhead{Object} &\colhead{Spectral} &\colhead{Class} &\colhead{RA} &\colhead{Dec} &\colhead{Binary$^1$} &\colhead{Comp$^2$} & \colhead{Distance} & \colhead{5 $\sigma$  Limit (Frame)}  & \colhead{5 $\sigma$  Limit (Coadd)} &\colhead{PA} &\colhead{Morpohlogy} \\
\colhead{} &\colhead{Type} &\colhead{} &\colhead{J2000} &\colhead{J2000} &\colhead{} &\colhead{} & \colhead{pc} & \colhead{Wm$^{-2}$s$^{-1}$Hz$^{-1}$} & \colhead{Wm$^{-2}$s$^{-1}$Hz$^{-1}$}&\colhead{E of N} &\colhead{}  &\colhead{Extent}  
 }
\startdata
CI Tau     & K7       & CTTS &  04$^{h}$33$^{m}$52$^{s}$.0 &  +22$^{o}$50$^{'}$30.2$^{"}$       &  N & N & 140     &2.15e-19 & 1.15e-19 & 35  &  Elongated   &  190 AU \\
RX J0338.2+1020& G9   & WTTS &  03$^{h}$38$^{m}$18$^{s}$.2 &  +10$^{o}$20$^{'}$17.1$^{"}$       &  N & N & 140     &1.57e-18 & 6.5e-19  & --  &  Symmetrical &  190 AU \\
LkCa 14    & M0     & WTTS &  04$^{h}$36$^{m}$19$^{s}$.1 &  +25$^{o}$42$^{'}$59.1$^{"}$       &  N & N & 140     &1.04e-19 & 7.97e-20 & 45  &  Elongated   &  120 AU \\
DI Cep     & G8   & CTTS &  22$^{h}$56$^{m}$11$^{s}$.5 &  +58$^{o}$40$^{'}$01.8$^{"}$       &  N & Y & 300     &3.18e-20 & 1.64e-20 & 85  &  Elongated   &  370 AU \\
\enddata
\tablenotetext{1}{Resolved close binary (within mask)}
\tablenotetext{2}{Faint companion}
\end{deluxetable}

\begin{deluxetable}{lllllllllllllllllllllll}
\tabletypesize{\tiny} \rotate \tablecaption{Photometry \label{phot}} 
\tablewidth{0pt} \tablehead{
\colhead{Object}     &\colhead{ U} & \colhead{V } &\colhead{ B }         &\colhead{ R }         & \colhead{I }         & \colhead{J   }       & \colhead{H  }       & \colhead{K }  &   \colhead{3.6  }   &   \colhead{4.5 }    &    \colhead{5.8 }   &       \colhead{8 }     &    \colhead{24   }   &    \colhead{12  }   &    \colhead{25  }   &     \colhead{60 }      &  \colhead{100 } &   \colhead{350 }   &     \colhead{450 }     &   \colhead{850 }     &  \colhead{1300} & \colhead{References} \\
\colhead{}
&\multicolumn{12}{c|}{mag}
&\multicolumn{9}{c}{mJy}
&\colhead{}} 
\startdata
LkCa 14    &  --       &       12.78&       11.51 &     10.81     &    --     &        9.34 &        8.71&        8.58&  8.47    &  8.54     &    8.49    &    8.44   &   --    &   --   &   --   &      --    &   --   &   --   &   --   &      --    &   --
& 1,2,6 \\
CI Tau     
&       14.27&       14.42&       13.25&     12.2   &        11.1&        9.47&        8.42&         7.8& 6.59      &  6.26      &   5.94     &     5.1    &   2.4   &   800     &  1380   &   2220  &  2254L     &   1725        &     846          &   324        &   190 
& 2,3,7,8,10  \\ 
DI Cep     
& --        &        11.9&       11.3 &  --        &      --       
&     9.3&         8.57&        7.95   & --       &    --   &   --      &    1000   &   --   &  1170      &  1540    &  4570L &  56960L  & -- &    103L        &     9L    	     & 	19L & 2,4,8,9,10  \\  
RXJ 0338   &           &    13.31  & 12.012 & 10.85 & 9.815  & 9.452  & 9.360  &   --     &   --   &   --  &   --     &   --   &   --   &      --    &   --
 &   --     &   --   &   --  &   -- & 2,5  \\
\enddata
\tablenotetext{1}{\citet{grankin08}}
\tablenotetext{2}{\citet{cat_2mass}}
\tablenotetext{3}{\citet{cat_xmmnewton}}
\tablenotetext{4}{\citet{ismailov07}}
\tablenotetext{5}{\citet{cat_usno}}
\tablenotetext{6}{\citet{hartmann05}}
\tablenotetext{7}{\citet{furlan06}}
\tablenotetext{8}{\citet{cat_iras}}
\tablenotetext{9}{\citet{cat_msx}}
\tablenotetext{10}{\citet{andrews05}}

\end{deluxetable}

\end{document}